\documentclass[prb,twocolumn,superscriptaddress,floatfix]{revtex4}
\usepackage{hyperref}
\usepackage{graphicx} 
\usepackage{amsmath}
\usepackage[dvips]{epsfig}

\begin{document}

\title{Rotating mixed $^3$He-$^4$He nanodroplets}

\author{Mart\'{\i} Pi}
\email{marti@fqa.ub.edu}
\affiliation{Departament FQA, Facultat de F\'{\i}sica,
Universitat de Barcelona. Diagonal 645,
08028 Barcelona, Spain}
\affiliation{Institute of Nanoscience and Nanotechnology (IN2UB),
Universitat de Barcelona, Barcelona, Spain.}

\author{Francesco Ancilotto}
\affiliation{Dipartimento di Fisica e Astronomia ``Galileo Galilei''
and CNISM, Universit\`a di Padova, via Marzolo 8, 35122 Padova, Italy}
\affiliation{ CNR-IOM Democritos, via Bonomea, 265 - 34136 Trieste, Italy }

\author{Jos\'e Mar\'{\i}a Escart\'{\i}n}
\affiliation{
Institut de Qu\'{\i}mica Te\`{o}rica i Computacional,
Universitat de Barcelona,
Carrer de Mart\'{\i} i Franqu\`{e}s 1,
08028 Barcelona, Spain.
}

\author{Ricardo Mayol}
\affiliation{Departament FQA, Facultat de F\'{\i}sica,
Universitat de Barcelona. Diagonal 645,
08028 Barcelona, Spain}
\affiliation{Institute of Nanoscience and Nanotechnology (IN2UB),
Universitat de Barcelona, Barcelona, Spain.}

\author{Manuel Barranco}
\affiliation{Departament FQA, Facultat de F\'{\i}sica,
Universitat de Barcelona. Diagonal 645,
08028 Barcelona, Spain}
\affiliation{Institute of Nanoscience and Nanotechnology (IN2UB),
Universitat de Barcelona, Barcelona, Spain.}
\affiliation{Universit\'e Toulouse 3, Laboratoire des Collisions, Agr\'egats et R\'eactivit\'e,
IRSAMC, 118 route de Narbonne, F-31062 Toulouse Cedex 09, France
}

\begin{abstract} 
Mixed $^3$He-$^4$He droplets created by hydrodynamic instability of a cryogenic fluid-jet 
may acquire angular  momentum during their passage through the nozzle of the experimental apparatus. 
These free-standing droplets cool down to very low temperatures undergoing isotopic segregation, developing a 
nearly pure $^3$He crust surrounding a very $^4$He-rich  superfluid  core. 
Here, the stability and appearance of rotating mixed helium nanodroplets
are investigated using Density Functional Theory 
for an isotopic composition that highlights, with some marked exceptions related to the existence of the
superfluid inner core, the 
analogies with viscous rotating droplets.
\end{abstract} 
\date{\today}

\maketitle

Ordinary liquids are known to form droplets held together by surface tension. When they are set into rotation, their spherical
shape experiences large 
deformations, evolving from oblate to prolate and 2-lobed, eventually fissioning if the rotational velocity is large enough.\cite{Bro80}
Experiments carried out by Plateau on olive oil droplets immersed in a mixture of water and alcohol with nearly the same
density, disclosed the sequence of droplet shapes as the angular velocity of the rotating shaft  to which they were attached 
increased.\cite{Pla63} The appearance and stability of, e.g., rotating celestial bodies,\cite{Cha65} atomic 
nuclei\cite{Coh74} and tektites,\cite{Bal15} to cite some quite different objects,  has been found to bear similarities with 
rotating classical droplets, adding an extrinsic interest to their study.  

Helium, in its two isotopes $^4$He and $^3$He,  is the only element in nature that may remain liquid and form  droplets 
at temperatures $(T)$ close to absolute zero.\cite{Toe04} Both isotopes may be  superfluid, with normal-to-superfluid transition temperatures
of 2.17 K ($^4$He) and 2.7 mK ($^3$He). At the temperatures of helium droplets experiments,
0.37 K for $^4$He\cite{Har95} and 0.15 K for $^3$He,\cite{Sar12} $^3$He is a normal fluid whereas $^4$He is superfluid. 
They constitute an ideal testground to study how superfluidity affects rotation, as they are isolated quantum systems
formed by atoms subject to the same bare interaction. 
 
Rotating superfluid $^4$He droplets made of $N_4= 10^8-10^{10}$ atoms, produced by hydrodynamic instability of a 
cryogenic fluid-jet, have been studied by coherent x-rays scattering from a free-electron laser,\cite{Gom14} revealing
 the presence of vortex lattices through the observation of Bragg patterns produced by Xe 
clusters captured by the vortex lines. Coherent diffractive imaging experiments 
 using extreme ultraviolet pulses have also been carried out,
aimed at providing informations about the droplet shapes.\cite{Lan18} 
Surprisingly, these studies have shown that 
superfluid $^4$He droplets follow the same shape sequence of rotating
viscous droplets made of normal fluid.
It has been shown that this is due to 
the presence of quantized vortices and capillary waves, 
whose interplay confers to the superfluid droplet the appearance of a classical 
rotating object.\cite{Anc18,Oco19} 
Until now, a deeper knowledge of  how superfluid droplets rotate has been hampered by the experimental difficulty of determining  their angular 
momentum, which is usually unknown.\cite{Oco19} 
This prevents a detailed comparison with theoretical models and the disclosure
of the precise quantum nature of such rotation.
Similar studies have been conducted very recently for rotating pure $^3$He droplets.\cite{QFC2019} 
These droplets are non-superfluid  and, as shown by Density Functional Theory (DFT) calculations,\cite{Pi19}
behave very much as classical rotating droplets. 
 
\begin{figure}[!]
\centerline{\includegraphics[width=1.0\linewidth,clip]{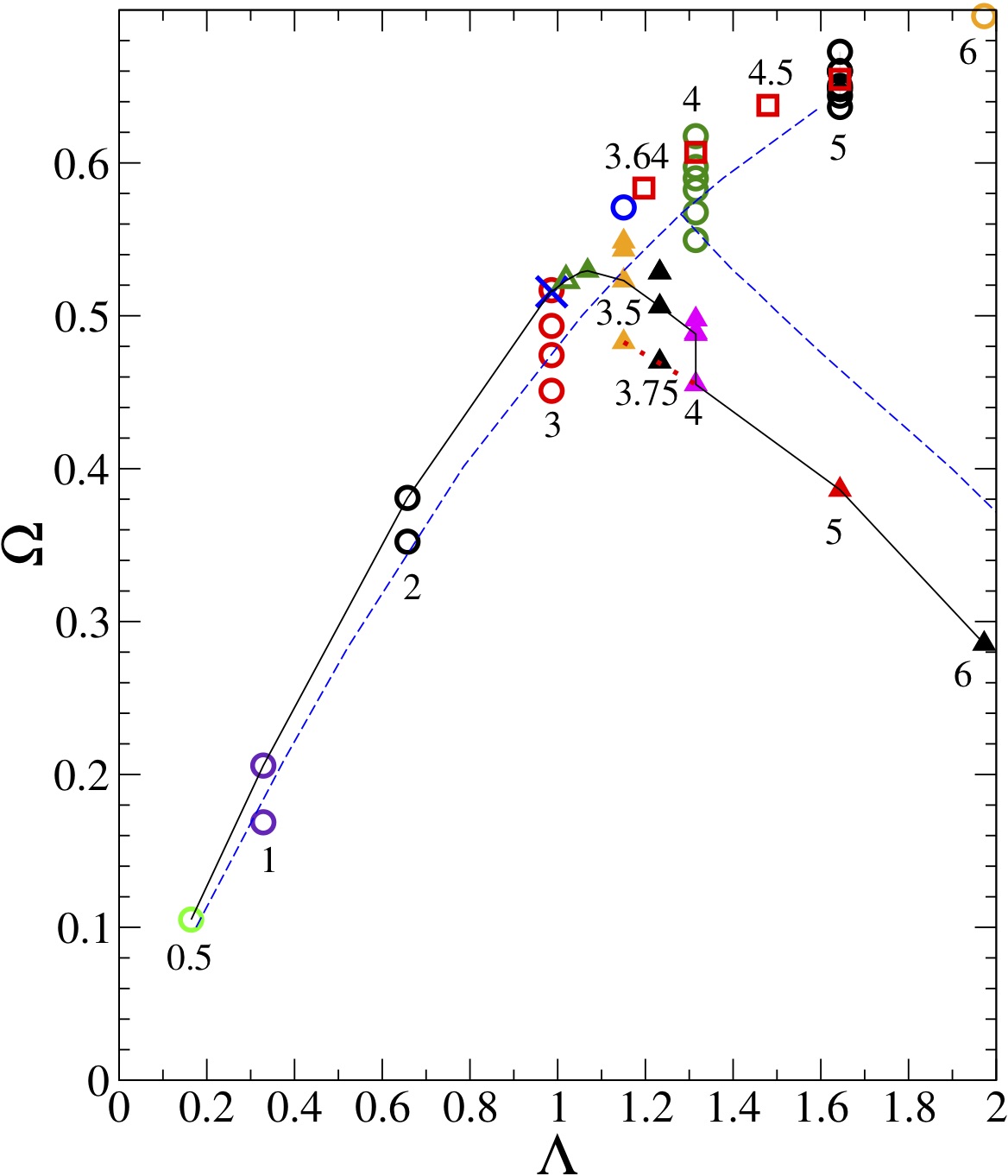}}
\caption{
Dimensionless angular velocity $\Omega$ vs.\ dimensionless angular momentum $\Lambda$. 
The numbers close to the symbols indicate the ${\cal L}$ value. 
Referring to the appearance of the outer surface of the $^3$He shell:
open circles, oblate configurations [$\Omega(\Lambda)$ rising branch]; 
triangles, prolate configurations
[$\Omega(\Lambda)$ falling branch]. 
Open squares,  3-lobed $^4$He  core configurations. 
The cross indicates  the oblate-to-prolate bifurcation point.
The solid black line --drawn as a guide to the eye-- connects the stable configurations for give ${\cal L}$,
and the dotted red line shows the region of metastable  fissioned  $^4$He core configurations. 
The dashed blue line is the DFT  result  for  pure $^3$He  droplets.\cite{Pi19}
}
\label{fig1}
\end{figure}

In liquid helium mixtures characterized by the $^3$He fraction $x_3= N_3/N$, with $N=N_3+N_4$, 
the normal-to-superfluid transition temperature decreases with increasing $x_3$.\cite{Edw92} At low $T$,
the mixture  undergoes a two-phase separation where a pure $^3$He phase coexists with a very $^4$He-rich 
mixture.\cite{Edw92} These properties are transferred to the
 mixed droplets, which at the experimental $T$ --sensibly that of pure $^3$He droplets\cite{Gre98}-- 
 experience a two-phase separation yielding a core-shell structure, with a
 crust made of $^3$He atoms in the normal state and a superfluid core mostly made of $^4$He atoms.\cite{Bar06}
 This segregation has been instrumental in finding the minimum number of $^4$He atoms needed to display 
 superfluidity.\cite{Gre98} 
 
The ability of forming self-bound isolated droplets made of a superfluid core enveloped by a normal fluid shell 
is  a unique characteristic of liquid helium mixtures at very low temperatures. Indeed,
Bose-Einstein condensates (BEC) immersed in a Fermi sea have been observed,\cite{Sch01} but these systems
are not self-bound, and only exist confined by an external trap to which the droplets adapt their shape;
self-bound  droplets made of  mixtures of bosonic cold gases have been also observed,\cite{Cab18,Sem18}
and self-bound Bose-Fermi droplets have  been studied theoretically.\cite{Rak19}  However,
these cold-gas mixtures  do not exist as phase-separated Bose-Fermi droplets, 
as they must remain in a mixed configuration to be self-bound.\cite{Pet15} 

The mixed normal fluid-superfluid structure of $^3$He-$^4$He droplets 
is expected to affect their structural properties as they are set into rotation.
A recent study addresses theoretically the classical rotation of droplets 
made of two immiscible viscous fluids.\cite{But20}   
When one component is superfluid, as in the case studied here,
then a quantum description is in order. We provide here such description.

\begin{figure}[!]
\centerline{\includegraphics[width=1.0\linewidth,clip]{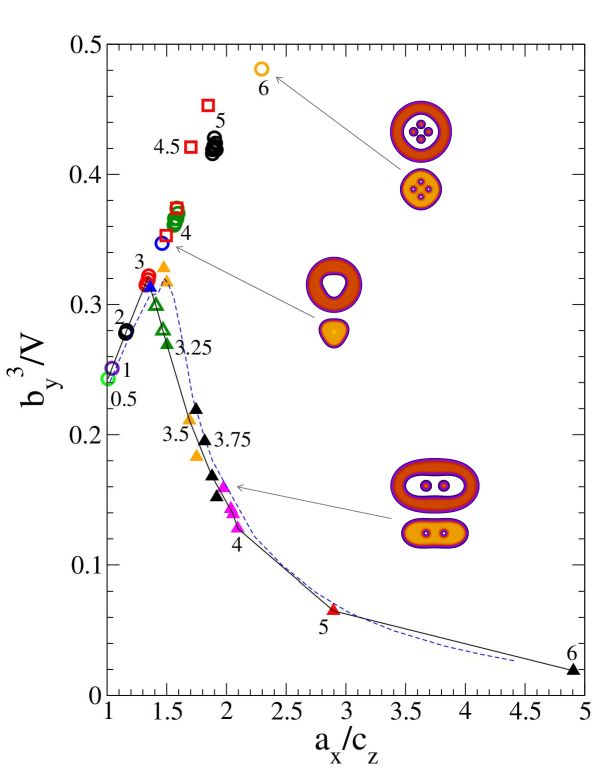}}
\caption{
Referring to the appearance of the outer surface of the $^3$He shell,  $b_y^3/V$ ratio vs. $a_x/c_z$, where $V$ is the volume of the
spherical droplet. 
Symbols  and lines have the same meaning as in Fig.~\ref{fig1}.
The numbers close to the symbols indicate the ${\cal L}$ value. 
The pictograms represent calculated $^3$He (top) and $^4$He (bottom) 2D densities for selected configurations.
}
\label{fig2}
\end{figure}
 
\section*{Results}

 We describe  the stability and shape transitions  of mixed He droplets within the DFT approach\cite{Bar06} taking as  case of study
 a $^4$He$_{1500}$-$^3$He$_{6000}$ nanodroplet ($x_3=80\%$). The size and composition of this nanodroplet,
which can be seen in Supplementary Fig.~1, 
have been chosen
(within the limitations imposed by the unavoidable computational cost of the calculations)
for physical reasons:
a thick  $^3$He crust is needed to model
a deformable container inside which the  superfluid $^4$He core  may undergo different shape transitions; a thin crust
would just adapt to the deforming core and  one should not expect a phenomenology much 
different from that of pure $^4$He droplets.\cite{Anc18}  
At the same time, the 
superfluid $^4$He inner droplet must be large enough to host 
a number of quantized vortices.\cite{Anc18} In superfluid liquid helium mixtures the vortex cores are filled with $^3$He and,  
depending on $x_3$, their radius can be up to 
five times larger than for pure $^4$He,\cite{Jez97}  
as shown in Supplementary Fig.~2. 
For the chosen $N_4$ value,  the number of vortices ($n_v$) is expected to be small.
    
 The calculations have been carried out as a function of the angular momentum per atom 
around the rotation $z$-axis,
 ${\cal L} = (L_3+L_4)/N = {\cal L}_3+{\cal L}_4$,   expressed in $\hbar$ units throughout this work. 
 For a given ${\cal L}$, the stable configuration is  that with the lowest energy including the rotational energy (Routhian).\cite{Bro80} 

 We characterize the appearance of the $^4$He core and $^3$He crust by the distance of their sharp surfaces
 (defined, for each isotope, by the locus at which the density equals that of the bulk 
liquid  divided by two\cite{Anc18,Pi19})  to 
 the center of mass of the droplet,   denoting  the distances along the $x, y$ and $z$ axes as $a_x$, $b_y$ and $c_z$, respectively. 
Dimensionless angular momentum $\Lambda$ and angular velocity $\Omega$ variables
 have been  defined  in Supplementary  Section 2. As in pure droplets,\cite{Bro80,Bal15,Oco19,But11}
 these variables are very useful to scale the results to droplets of different  size for a given composition.\cite{But20} 
 
\begin{figure}[!]
\centerline{\includegraphics[width=1.0\linewidth,clip]{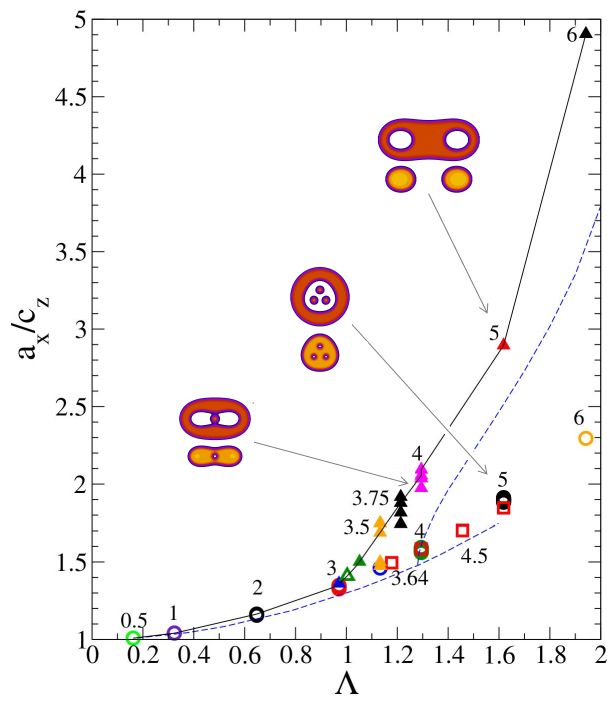}}
\caption{
$a_x/c_z$ ratio of the outer surface of the $^3$He shell vs $\Lambda$. 
 Symbols and lines have the same meaning as in Fig.~\ref{fig1}.
 The numbers close to the symbols indicate the ${\cal L}$ value.
The pictograms represent calculated $^3$He (top) and $^4$He (bottom) 2D DFT densities for selected configurations.
}
\label{fig3}
\end{figure}

\begin{figure*}[!]
\centerline{
\includegraphics[width=1.0\linewidth,clip]{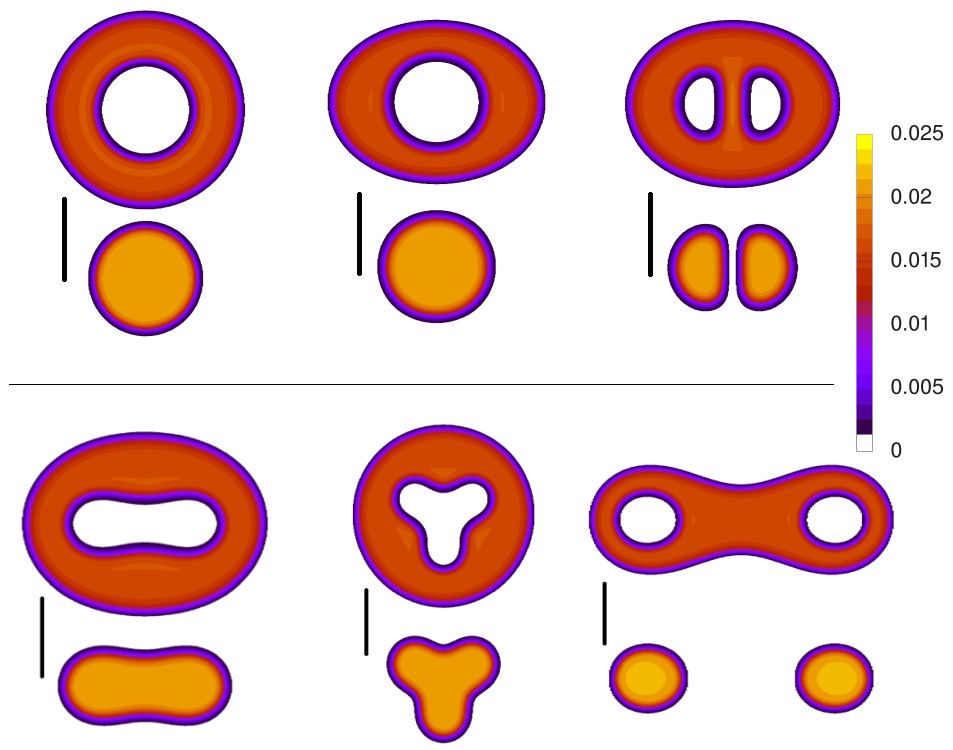}
}
\caption{
Top: Density of the $^4$He$_{1500}$-$^3$He$_{6000}$ nano droplet 
on a symmetry plane containing the rotational axis.
From left to right, spherical configuration at ${\cal L}=0$; oblate vortex-free $^4$He core configuration at ${\cal L}=3$, and oblate one-vortex 
configuration with quantum circulation $m=1$
at ${\cal L}=3$.  
Bottom :
Density of the $^4$He$_{1500}$-$^3$He$_{6000}$ nanodroplet on the symmetry plane perpendicular to the rotational axis.
From left to right, prolate vortex-free $^4$He core configuration at ${\cal L}=3.5$; oblate 3-lobed $^4$He core configuration at ${\cal L}=4$,
and prolate fissioned $^4$He core configuration at ${\cal L}=6$. 
For each configuration, the top density corresponds to the $^3$He crust and the bottom density to the $^4$He core. 
The black vertical bars represent a distance of 40\,\AA. 
The color bar shows the atom density in units of \AA$^{-3}$ and is common to all 
configurations.
}
\label{fig4}
\end{figure*}

The detailed energetics and morphologic characteristics of rotating mixed helium nanodroplets 
are collected in   Figs.~\ref{fig1}-\ref{fig4} and Supplementary Table 1.
Figure~\ref{fig1} shows the stability diagram  
and constitutes the main result of this work.
Figure~\ref{fig2} provides
information on the shapes of the droplets through relationships between the
geometrical parameters  $a_x$, $b_y$, and $c_z$ that
characterize the shape of the outer $^3$He surface.\cite{Bal15,Lan18}
Lastly, Fig.~\ref{fig3} connects the shapes of the droplets
with the dimensionless angular momentum $\Lambda$.

Our study has unveiled  a rich variety of stable and metastable configurations.
 As $\Lambda $ increases from zero, the $^3$He crust  
 becomes oblate as it happens in normal fluids.\cite{Bro80} At variance,
 the  superfluid $^4$He core remains spherical, becoming axisymmetric 
 only when the angular momentum of the droplet is large enough. Since  the  
 superfluid $^4$He core cannot be set  into rotation around the symmetry axis because 
 it is quantum-mechanically forbidden, all the angular momentum is stored in the $^3$He crust. 
This is in stark contrast with the case of pure $^4$He droplets, where oblate configurations 
can exist because they host 
quantized vortices.\cite{Anc15,Anc18,Oco19}  In rotating mixed droplets, the $^4$He core may remain axisymmetric because
  the angular momentum is mainly stored in the  $^3$He shell, which acts as a rotating deformable container.
  We have found that this happens 
 up to the oblate-to-prolate bifurcation point at $(\Lambda,\Omega)=(0.99, 0.52)$, as shown in Fig.~\ref{fig1}.
  Thus,  for this composition, oblate stable configurations are vortex-free
  for droplets of the size studied here.
We recall that for pure $^3$He droplets the bifurcation point is at
 $(\Lambda, \Omega)=(1.28, 0.57)$.\cite{Pi19} 
 
 In the prolate branch, the outer surface of the $^3$He  crust is triaxial ellipsoid-like up to $\Lambda \sim 1.64$,
 where it becomes 2-lobed.  At variance, the superfluid
 $^4$He core becomes 2-lobed at $\Lambda \sim1.05$, i.e. immediately after bifurcation.
 This is due to the small surface tension of the $^3$He-$^4$He interface, 0.016 K\AA$^{-2}$, as compared to 
 that of the $^3$He free surface, 0.113 K\AA$^{-2}$.  Pure $^3$He droplets become 2-lobed at $\Lambda=1.85$.\cite{Pi19}
 
 Prolate stable configurations with simply connected $^4$He cores have been  found only  in a narrow angular momentum
  range $3.0 \leq {\cal L} \leq 3.5$, where  the core shape evolves from spheroidal to triaxial to 2-lobed. 
Due again to the small surface tension of the $^3$He-$^4$He interface, 
the $^4$He core undergoes fission when the $^3$He crust is still 
 triaxial ellipsoid-like. The resulting  stable prolate configuration  consist of a fissioned
$^4$He core  inside a rotating triaxial $^3$He crust. The transition from simply connected to fissioned
 $^4$He core configurations  appears as a jump in the $\Omega(\Lambda)$ curve on Fig.~\ref{fig1} 
at $\Lambda = 1.31$. Notice that the jump to fissioned inner $^4$He core
(and the associated hysteresis loop as well) could not be predicted out of simple models based on
energy minimization restricted to simply connected shapes.
 
We have looked for prolate configurations with a fissioned $^4$He core 
 hosting a vortex in each moiety. After phase-imprinting them,\cite{Anc18} 
vortices are eventually expelled in the course of the numerical relaxation; 
 we conclude that   these configurations are not stable  for up to the largest  
$\Lambda$ addressed in this study,  $\Lambda = 1.97$, a fairly large value for current experiments.

In spite of the fact that the lowest-energy configurations of the inner $^4$He nanodroplet are  mostly vortex-free
(most of the angular momentum being efficiently stored in the $^3$He shell),
we have found a number of vortex-hosting  \textit{metastable} configurations  in our calculations.   
Remarkably,  along  the oblate branch (in the metastable region beyond the bifurcation point)  
we have found that  configurations with $n_v=1$ have lower energy than vortex-free configurations.
In this region,  configurations hosting up to 4 vortices appear  at 
${\cal L>}$ 3.5. All these configurations  are  metastable, as they may decay
to prolate configurations lying at much lower energy (see 
Supplementary Table 1 for details).

In the oblate branch, we  have also addressed multiply-charged quantum vortices with charge (quantum circulation)  $m=2-4$, 
and their relative stability with respect to configurations with $n_v=2-4$ and $m=1$.  Multiply charged vortices in 
BEC have been found that are stabilized temporarily by the external confining potential.\cite{Shi04,Oka07} In mixed helium droplets,
the self-bound thick $^3$He shell yields a confining potential that plays the same role.
We have found (see Supplementary Table 1) that, depending on ${\cal L}$,  multiply charged single vortex configurations with
 charge $m$  are more stable than $n_v=m$ 
singly-charged vortices, and hence cannot decay into them as it would happen 
in pure $^4$He. 
This is likely due
to the presence of $^3$He in the expanded vortex cores and the $^3$He crust, which together define
a region similar to the rotating annulus used
to study quantized superfluid states and vortices in the first experiments on 
quantized circulation in superfluid liquid $^4$He.\cite{Vin61,Don91}
 
 Along the prolate branch, metastable $^4$He configurations with $n_v=1-2$  have been found 
 (see Supplementary Table 1)  where the angular momentum of the $^4$He 
 core is shared between vortices and capillary waves as shown in Supplementary Fig.~3. This is
 similar to what happens  in   spinning $^4$He droplets.\cite{Anc18,Oco19}
 When the $^4$He core becomes 2-lobed,  the neck  connecting   them gets thinner as ${\cal L}$ increases and 
 eventually  the most stable configuration is the fissioned one.
 
 Three-lobed configurations were predicted to appear in classical droplets rotating at high velocities.\cite{Bro80}
These configurations are metastable with respect to prolate 2-lobed configurations and were not expected to be accessible 
experimentaly. However,  3-lobed configurations  were
obtained,\cite{Ohs00} stabilized by forcing the droplet  into large amplitude
periodic oscillations --thus not in gyrostatic equilibrium.
Charged $^4$He drops magnetically levitated have been studied displaying  $\ell$ =2-4 oscillation modes induced 
by   a rotating deformation of the droplet.\cite{Whi98}
More recently,  triangular-shaped magnetically levitated water  droplets have been found\cite{Hil08} where
 the amplitude of the surface oscillation is small and the equilibrium shape could be observed clearly  for about 100 revolutions. 
 
 In the water droplet experiments, some surfactant was added to water to decrease the surface tension, helping the 3-lobed configurations 
 to be formed.  In the case of mixed helium droplets, the surface tension of the $^3$He-$^4$He interface is already small, hence it is
possible to have metastable 3-lobed $^4$He core configurations while the outer surface 
of the $^3$He crust is still oblate. Indeed, we have found such configurations in the $3.64 \leq {\cal L} \leq 5$ range,
one of which is shown in Fig.~\ref{fig4} (see also the central pictogram in Fig.~\ref{fig2}). 
The 3-lobed  bifurcation sets in at $(\Lambda, \Omega)= (1.20, 0.58)$.
We have looked for metastable 4-lobed configurations but have not found any,
the droplet always decaying into fissioned $^4$He 
prolate configurations.

\section*{Discussion}

The presence of a superfluid $^4$He core inside a normal fluid $^3$He 
droplet produces remarkable changes in its rotational
properties. Oblate configurations are affected as the superfluid 
core cannot participate in the rotation because of its symmetry
around the rotational axis. At variance with rotating pure $^4$He nanodroplets, 
stable vortex-hosting oblate configurations are absent as it 
is energetically more favorable for the droplet to store  angular 
momentum in the deformable $^3$He crust than in the superfluid $^4$He core. 
Vortex-hosting configurations are  absent as equilibrium configurations in the 
prolate branch as well. While this could be due to the nanoscopic size of 
the droplets studied here, 
we want to stress that, as Supplementary Table 1 shows, the energy difference between vortex-hosting and vortex-free
configurations can be very small  and likely separated by small energy barriers, so 
one should not discard that both types of configurations are found in actual experiments.

As for pure $^4$He droplets, the presence of vortices  can be experimentally tested by doping 
the droplets with heliophilic impurities.\cite{Gom14,Oco19}
 These impurities are captured by the droplet and  sink into the $^4$He core.\cite{Bar06} 
Coherent x-ray scattering would reveal the space distribution of the impurities, which might
arrange along the vortex cores if vortex arrays are present,\cite{Gom14,Oco19} 
producing  otherwise interference patterns very different from those of vortex-hosting droplets.
 
 In the prolate branch, due to the small surface tension of the $^3$He-$^4$He interface,
the $^4$He core fissions already at moderate angular velocities  
and superfluid, multiple  cores configurations are the more stable ones.   
 Diffractive imaging of multiple 
 connected $^4$He cores for droplets of the size addressed in this
 work is challenging at present  because they are small and the contrast is expected to be small, but likely not for 
 the $N=10^8-10^{11}$ droplets in  ongoing experiments. 

It is less obvious whether  metastable 3-lobed $^4$He core  configurations may be experimentally detected; in the case of
classical droplets, they were identified\cite{Ohs00,Hil08} twenty years after being predicted.\cite{Bro80} 
These configurations are highly unstable; for instance, Supplementary Table 1 shows that a clear 3-lobed configuration
as that at ${\cal L}=4$ can either decay to the metastable oblate $n_v=1$ vortex 
configuration, $\sim 8$\;K below it, or to  the  stable prolate, fissioned core configuration, about $37$\;K below it.

An investigation on viscous immiscible two-fluid droplets has been conducted recently\cite{But20} that complements the present study on
droplets made of two quantum fluids of limited solubility.  Last but not least,
our results  can be used as benchmark for the applicability of these classical calculations to the quantum domain.

\section*{Methods}

We have considered in the present work mixed helium droplets 
at ``zero temperature'', this meaning a temperature 
so low (a few mK) that
thermal effects on the energetics and morphology of the droplet are negligible, 
$^3$He is in the normal phase, and $^4$He is superfluid.

The droplet is described within the DFT approach\cite{Anc17} using the density functional
of Ref. \onlinecite{Bar97}. Due to the large number of $^3$He atoms in this study, $^3$He 
can be treated  semiclassically in the Thomas-Fermi approximation.\cite{Pi19} 
The DFT  equations have been solved adapting the 4He-DFT BCN-TLS computing package\cite{Pi17} to the case of helium mixtures. 
Further details are given in Supplementary Section 1.
 
\section*{Code availability}
The results have been obtained adapting  the 4He-DFT BCN-TLS computing package which is freely available.\cite{Pi17}

\begin{acknowledgments}
We are most indebted to  Andrey Vilesov for informations  on their 
ongoing experiments with mixed helium droplets that have 
motivated  and clarified some aspects of this work,  and to Sam Butler for useful exchanges. This work has been 
performed under Grant No  FIS2017-87801-P (AEI/FEDER, UE).
J.M.E. acknowledges support from Ministerio de Ciencia e Innovaci\'{o}n of Spain
through the Unidades de Excelencia ``Mar\'{\i}a de Maeztu'' grant MDM-2017-0767.
M.B. thanks the Universit\'e F\'ed\'erale Toulouse Midi-Pyr\'en\'ees for financial support  throughout 
the ``Chaires d'Attractivit\'e 2014''  Programme IMDYNHE.
\end{acknowledgments}

\newpage

{\bf Supplementary Information for ``Rotating  mixed $^3$He-$^4$He nanodroplets''}

\bigskip

\section*{S1.- The DFT approach for mixed $^3$He-$^4$He nanodroplets}

 The DFT equations obtained by functional variation of the energy density
are formulated in a rotating frame of reference with constant angular velocity $\omega$ around the $z$ axis.\cite{Anc18} 
In terms of the DFT Hamiltonians ${\cal H}_3[\rho_3,\rho_4]$ and ${\cal H}_4[\rho_3,\rho_4]$,\cite{Bar97}
\begin{equation}
\begin{aligned}
&\left\{{\cal H}_3[\rho_3,\rho_4] \,- \frac{m_3}{2}\,\omega^2 (x^2+y^2)\right\} \,\Psi_3(\mathbf{r})  =  \,\mu_3 \, \Psi_3(\mathbf{r}) 
\\
&\left\{{\cal H}_4[\rho_3,\rho_4] \,-\omega \hat{L}_4\right\} \,\Psi_4(\mathbf{r})  =  \,\mu_4 \, \Psi_4(\mathbf{r})
\label{eq1}
\end{aligned}
\end{equation}
where $\mu_3$($\mu_4$) is the $^3$He($^4$He) chemical potential, $\hat{L}_4$ is the $^4$He angular momentum operator,
 and $\Psi_3(\mathbf{r})$  and  $\Psi_4(\mathbf{r})$ are the
real $^3$He and complex $^4$He  effective  wavefunctions 
related to the atom densities as  $\Psi_3^2(\mathbf{r})=\rho_3(\mathbf{r})$ and 
$|\Psi_4(\mathbf{r})|^2=\rho_4(\mathbf{r})$. These equations are
solved imposing a given value of the total angular momentum per atom, which requires finding iteratively the value of $\omega$. 
Classically, this corresponds to torque-free droplets with an initially prescribed rotation, as they are isolated.

Vortices are nucleated  in the $^4$He core using  the imprinting  procedure\cite{Anc18} by which  $n_v$
vortex lines parallel to the $z$ axis are initially created, \textit{i.e.}, one starts  the  iterative solution of equation (\ref{eq1}) from 
the wave function
\begin{equation}
\Psi_4(\mathbf{r})=\sqrt{\tilde\rho_4(\mathbf{r})}\, 
\prod _{j=1}^{n_v} {(x-x_j)+\imath (y-y_j) \over \sqrt{(x-x_j)^2+(y-y_j)^2}}
\label{eq2}
\end{equation}
where   $(x_j, y_j)$ is the initial position of the $j$-vortex linear  core with
respect to the $z$-axis of the droplet, and $\tilde\rho_4(\mathbf{r})$ is the vortex-free 
$^4$He density.
The initial vortex positions are guessed and
during the functional minimization
of the total energy, both the vortex positions and droplet density
are allowed to change
to provide at convergence the lowest energy
configuration for the chosen value  of $L_z$.

\section*{S2.- Scaled angular momentum and angular velocity}

Classical rotating droplets subject to surface tension and centrifugal forces are characterized by two dimensionless variables,
angular momentum $\Lambda$ and velocity $\Omega$ that allow to describe the sequence of droplet shapes in a universal 
phase diagram, independently of the droplet size.\cite{Bro80,But11} These variables have also been used  to 
characterize pure $^4$He and $^3$He droplets within the DFT approach.\cite{Anc18,Pi19} 

The definitions  of  $\Lambda$ and  $\Omega$ have been generalized to the case of droplets made of two
immiscible fluids  as follows.\cite{But20}
An effective particle density $\rho_{\rm eff}$ and  mass $m_{\rm eff}$ are introduced such that the moment of inertia of a
spherical droplet with such effective mass and density  coincides with that of 
a spherical droplet of inner radius $R_i$ and atom density $\rho_4$ 
surrounded by a spherical shell of outer radius $R_o$ and atom density $\rho_3$:
\begin{equation}
m_{\rm eff} \rho_{\rm eff} =  \frac{ m_4 \rho_4 R_i ^5 + m_3 \rho_3 (R_o^5 - R_i^5) }{R_o^5} 
\label{eq10}
\end{equation}
together with an effective surface tension
\begin{equation}
\gamma_{\rm eff} = \frac{\gamma_{34} R_i + \gamma_3 R_o}{R_o}
\label{eq11}
\end{equation}
where $\gamma_{34}$ is the surface tension of the $^3$He-$^4$He interface and $\gamma_3$ that of the $^3$He free surface.
The scaled variables $\Lambda$ and $\Omega$  are then 
written as for pure droplets \cite{Bro80,But11} but replacing $m$ and $\gamma$
with $m_{\rm eff}$ and $\gamma_{\rm eff}$, respectively: 
\begin{equation}
\Lambda = \frac{ N \hbar}{\sqrt{8 \,\gamma_{\rm eff}\, R_o^7 \, m_{\rm eff} \, \rho_{\rm eff} }}\; {\cal L}
\label{eq12}
\end{equation}
\begin{equation}
\Omega= \sqrt{\frac{m_{\rm eff}\, \rho_{\rm eff} \, R_o^3}{8\, \gamma_{\rm eff}}} \; \omega \;.
\label{eq13}
\end{equation}
Note that $\Lambda$ is written in terms of the total angular momentum per atom in $\hbar$ units, 
${\cal L} = (L_3+L_4)/(N_3+N_4) = L/N$.

The values of the magnitudes entering the above equations are: $\gamma_3= 0.113$\;K\;\AA$^{-2}$, 
$\gamma_{34}= 0.016$\;K\;\AA$^{-2}$, $\hbar^2/(2m_3) = 8.0418$\;K\;\AA$^2$, $\hbar^2/(2m_4) = 6.0597$\;K\;\AA$^2$,
$\rho_3 = 0.016347$\;\AA$^{-3}$, and $\rho_4 = 0.021836$\;\AA$^{-3}$. For the $N_3=6000$, $N_4=1500$ droplet, one
gets $R_i= 25.40$\;\AA{} and $R_o= 47.02$\;\AA{}, and hence $\Lambda = {\cal L}/3.042$ and $\Omega = 10.61 \; \hbar \omega$, if 
$\hbar \omega$ is given in K.

\begin{table*}[!]
\begin{tabular}{|c|ccccccc|ccccc|ccc|}
\hline
\hline 
  &$\Lambda$  & $\Omega$ & ${\cal L} $ & ${\cal L}_3 $ &${\cal L}_4 $ &$ \hbar \omega$& ${\cal R}$ &
 \multicolumn{5}{c|}{$^3$He} & \multicolumn{3}{c|}{$^4$He} \\  \cline{9-16} 
 & & &($\hbar$) & ($\hbar$)&($\hbar$) & ($\times 10^{-2}$K)& (K) & 
   $a_x$ (\AA)& $b_y$ (\AA) & $c_z$ (\AA) & $a_x/c_z$ & $b_y^3/V$& $a_x$ (\AA)& $b_y$ (\AA) & $c_z$ (\AA) \\
\hline
\hline
{\bf \underline{O}}  &0.164 & 0.105   & 0.5  & 0.50 & 0.00 & 0.99 & -22488.5 & 47.36 & 47.36 & 46.95 & 1.009& 0.243& 25.29 & 25.26 & 25.35 \\
{\bf \underline{O}}   & 0.329&  0.206 & 1    & 1.00 & 0.00 & 1.94 & -22433.3 & 47.86 & 47.86 & 45.96 &1.041 &0.251 & 25.24 & 25.22 & 25.43 \\
O1v  & 0.329& 0.169 & 1    & 0.80 & 0.20 & 1.59 & -22385.4 & 47.87 & 47.87 & 45.93 &1.042 &0.251& 29.15 & 29.15 & -- \\
{\bf \underline{O}}  &0.657 &  0.381  & 2    & 2.00 & 0.00 & 3.59 & -22223.4 & 49.64 & 49.63 & 42.65 & 1.164&0.280 & 25.26 & 25.19 & 25.31 \\
O1v  &0.657 & 0.352 & 2    & 1.80 & 0.20 & 3.32 & -22198.8 & 49.49 & 49.49 & 42.81 &1.156 &0.278 & 28.90 & 28.90 & -- \\
{\bf \underline{O}}   & 0.986& 0.517  & 3    & 3.00 & 0.00 & 4.87 & -21903.87  & 52.02 & 52.02 & 38.57 &1.349 & 0.322& 25.57 & 25.57 & 24.31 \\
O1v  &0.986 & 0.493 & 3    & 2.80 & 0.20 & 4.65 & -21897.4 & 51.83 & 51.83 & 38.57 &1.344 &0.319 & 28.67 & 28.67 & -- \\
O1v2m & 0.986& 0.474& 3    & 2.60 & 0.40 & 4.47 & -21875.2 & 51.67 & 51.67 & 38.77 &1.333 &0.316 & 32.59 & 32.59 & -- \\
O1v3m & 0.986&0.451 & 3    & 2.40 & 0.60 & 4.25 & -21851.2 & 51.61 & 51.61 & 38.90 &1.327 &0.315 & 35.77 & 35.77 & -- \\
O   &  1.151& 0.571& 3.5   & 3.50 & 0.00 & 5.38 & -21711.4 & 53.33 & 53.33 & 36.50 &1.461 &0.347 & 25.89 & 25.89 & 23.47 \\
O1v &  1.151& 0.550& 3.5   & 3.50 & 0.20 & 5.18 & -21713.0 & 53.12 & 53.12 & 36.33 &1.462 &0.343 & 28.67 & 28.67 & -- \\
O1v2m& 1.151& 0.533& 3.5   & 3.10 & 0.40 & 5.02 & -21697.0 & 52.93 & 52.93 & 36.51 &1.449 &0.339 & 32.48 & 32.48 & -- \\
O     &1.315 & 0.618& 4    & 4.00 & 0.00 & 5.82 & -21501.1  & 54.66 & 54.66 & 34.48 &1.585 &0.374 & 26.36 & 26.36 & 22.47 \\
O1v   & 1.315& 0.597& 4    & 3.80 & 0.20 & 5.63 & -21510.2  & 54.46 & 54.46 & 34.17 &1.594 &0.370 & 28.80 & 28.80 & -- \\
O1v2m & 1.315& 0.582& 4    & 3.60 & 0.40 & 5.49 & -21499.8  & 54.27 & 54.27 & 34.26 &1.584 &0.366 & 32.42 & 32.42 & -- \\
O1v3m & 1.315& 0.568& 4    & 3.40 & 0.60 & 5.35 & -21488.6  & 54.10 & 54.10 & 34.43 & 1.571&0.363 & 35.73 & 35.73 & -- \\
O1v4m & 1.315& 0.550& 4    & 3.20 & 0.80 & 5.18 & -21474.8  & 54.01 & 54.01 & 34.59 &1.561 & 0.361& 38.38 & 38.38 & -- \\
O2v   &1.315 & 0.589& 4    & 3.70 & 0.30 & 5.56 & -21496.8  & 54.18 & 54.26 & 34.62 &1.565 &0.366 & 30.87 & 29.23 & 18.72 \\
O1v   &1.644 &0.673 & 5    & 4.80 & 0.20 & 6.34 & -21059.9  & 57.17 & 57.17 & 30.10 &1.899 & 0.428& 29.42 & 29.42 & -- \\
O1v2m &1.644 &0.660 & 5  & 4.60 & 0.40 & 6.22 & -21059.4  & 56.99 & 56.99 & 29.87 &1.908 & 0.424& 32.56 & 32.56 & -- \\
O1v3m & 1.644& 0.648& 5    & 4.40 & 0.60 & 6.11 & -21057.0  & 56.81 & 56.81 & 29.93 &1.898 &0.420 &  35.76 & 35.76 & -- \\
O1v4m & 1.644& 0.637& 5    & 4.20 & 0.80 & 6.00 & -21053.5  & 56.65 & 56.65 & 30.08 & 1.883&0.416 & 38.55 & 38.55 & -- \\
O2v   &1.644 &0.660 & 5  & 4.63 & 0.37 & 6.22 & -21057.7  & 57.27 & 56.76 & 29.94 &1.913 & 0.419& 33.73 & 30.13 & -- \\
O3v   & 1.644& 0.650& 5     & 4.50 & 0.50 & 6.13 & -21050.2  & 56.88 & 56.89 & 30.01 &1.895 & 0.422& 34.36 & 34.30 & 3.05  \\
O4v  & 1.644& 0.644& 5    & 4.39 & 0.61 & 6.07 & -21039.6  & 56.77 & 56.77 & 30.14 & 1.884&0.419 & 36.47 & 36.47 & 11.80 \\
O4v  & 1.972& 0.696& 6    & 5.39 & 0.71 & 6.56 & -20565.5  & 59.44 & 59.44 & 25.90 & 2.295&0.481 & 38.13 & 38.13 &   7.34 \\
\hline
3L   &1.197 & 0.584& 3.64 & 3.628 & 0.008 & 5.50 & -21655.7 & 53.74 & 53.62 & 35.94 &1.495 &0.353 & -- & -- & -- \\
3L   &1.315 & 0.607& 4    & 3.880 & 0.120 & 5.72 & -21502.4 & 54.44 & 54.68 & 34.37 &1.584 &0.374 & -- & -- & -- \\
3L   &1.479 & 0.638& 4.5  & 4.278 & 0.222 & 6.01 & -21282.9 & 54.79 & 56.85 & 32.20 &1.702 &0.421 & -- & -- & -- \\
3L   &1.644 & 0.655& 5    & 4.541 & 0.459 & 6.17 & -21056.0 & 55.63 & 58.28 & 30.12 &1.847 &0.453 & -- & -- & -- \\
\hline
P & 0.986& 0.516& 3   & 2.994 & 6.3$\times 10^{-3}$ & 4.86 & -21903.86 & 52.51 & 51.54   & 38.57 & 1.361& 0.313& 27.83 & 23.36 & 24.22 \\
{\bf \underline{P}} & 1.019& 0.523& 3.1 & 3.060 & 0.040  & 4.93 & -21867.1 & 53.78 & 50.72 & 38.13 & 1.410& 0.299& 31.38 & 20.20 & 23.62 \\
{\bf \underline{P}} &1.052 & 0.528& 3.2 & 3.116 & 0.084  & 4.98 & -21829.9 & 55.34 & 49.64 & 37.69 & 1.468& 0.280& 34.17 & 18.05 & 22.85 \\
{\bf \underline{P}} & 1.068& 0.529& 3.25 & 3.141 & 0.109  & 4.99 & -21811.2 & 56.27 & 48.96 & 37.47 & 1.502& 0.269& 35.51 & 17.18 & 22.39 \\
{\bf \underline{P}} &1.151 & 0.523& 3.5 & 3.264 & 0.236  & 4.93 & -21718.0 & 61.37 & 45.20 & 36.31 & 1.690& 0.211& 41.44 & 14.46 & 19.94 \\
P1v & 1.151& 0.549& 3.5& 3.299 & 0.201 & 5.17 & -21712.9 & 54.52 & 51.73 & 36.34 & 1.500&0.317 & 29.39 & 27.97 & --   \\
P2v & 1.151& 0.544& 3.5& 3.239 & 0.261 & 5.12 & -21697.7 & 53.78 & 52.31 & 36.47 & 1.475&0.328 & 30.95 & 29.93 & --  \\
Pf  & 1.151& 0.483& 3.5& 2.897 & 0.603 & 4.55 & -21705.5 & 63.24 & 43.04 & 36.14 & 1.750&0.183 & -- & -- & --   \\
{\bf \underline{P}} & 1.233& 0.506& 3.75 & 3.385 & 0.365 & 4.77 & -21627.0 & 66.12 & 41.90 & 35.18 & 1.879&0.168 & 46.69 & 12.94 & 17.47   \\
P1v & 1.233&0.528& 3.75 & 3.488 & 0.262 & 4.98 & -21617.4 & 63.96 & 44.00 & 35.20 & 1.817&0.195 & 37.43 & 22.87 & --   \\
P2v & 1.233&0.528& 3.75 & 3.329 & 0.421 & 4.98 & -21603.1 & 61.76 & 45.70 & 35.40 & 1.745&0.219 & 41.05 & 23.98 & 14.27   \\
Pf  & 1.233&0.470& 3.75 & 3.003 & 0.747 & 4.43 & -21621.6 & 67.08 & 40.53 & 34.98 & 1.918&0.152 & 52.74 & -- & --   \\
P    & 1.315& 0.488& 4 & 3.499 & 0.501 & 4.60 & -21539.1 & 70.04 & 39.35 & 34.08 & 2.055&0.139 & 51.45 & 11.65 & 15.04   \\
P1v  & 1.315& 0.492& 4 & 3.422 & 0.578 & 4.64 & -21527.8 & 69.08 & 40.00 & 34.13 & 2.024&0.091 & 51.06 & 13.76 &  --  \\
P2v  & 1.315& 0.498& 4 & 3.389 & 0.611 & 4.69 & -21512.3 & 67.75 & 41.10 & 34.31 &1.975 &0.159 & 49.41 & 19.45 & 11.31   \\
{\bf \underline{Pf}}   &1.315 & 0.455& 4 & 3.115 & 0.885 & 4.29 & -21539.8 & 70.77 & 38.20 & 33.80 &2.094 &0.128 & -- & -- & --   \\
{\bf \underline{Pf}}   & 1.644& 0.386& 5 & 3.640 & 1.360 & 3.64 & -21242.8 & 83.55 & 30.56 & 28.86 &2.895 &0.065 &-- & -- & --  \\
{\bf \underline{Pf}}   & 1.972& 0.385& 6 & 4.317 & 1.683 & 2.69 & -20999.7 & 97.61 & 20.19 & 19.90 &4.905 &0.019 &-- & -- & --  \\
\hline
\hline
\end{tabular}
\caption{Characteristics of the  rotating $^3$He$_{6000}$-$^4$He$_{1500}$ nanodroplet ($x_3=80\%$). O: oblate configurations; 
P: prolate configurations; 3L:  3-lobed $^4$He core configurations (the outer surface of the $^3$He crust 
is sensibly oblate); Pf: fissioned $^4$He core configurations. 
``1v3m'' e.g., means 1 vortex with circulation (charge) 3.
${\cal L} = {\cal L}_3 + {\cal L}_4 = L_3/N +L_4/N$ with $N=N_3+N_4$, and $V$ is the volume of the spherical droplet at rest.
The most stable configuration for each ${\cal L}$ value is underlined boldface.
Some entries are empty because they are not well defined or meaningless.\\ 
\label{Table1}
}
\end{table*}

\begin{figure*}[!]
\centerline{\includegraphics[width=0.8\linewidth,clip]{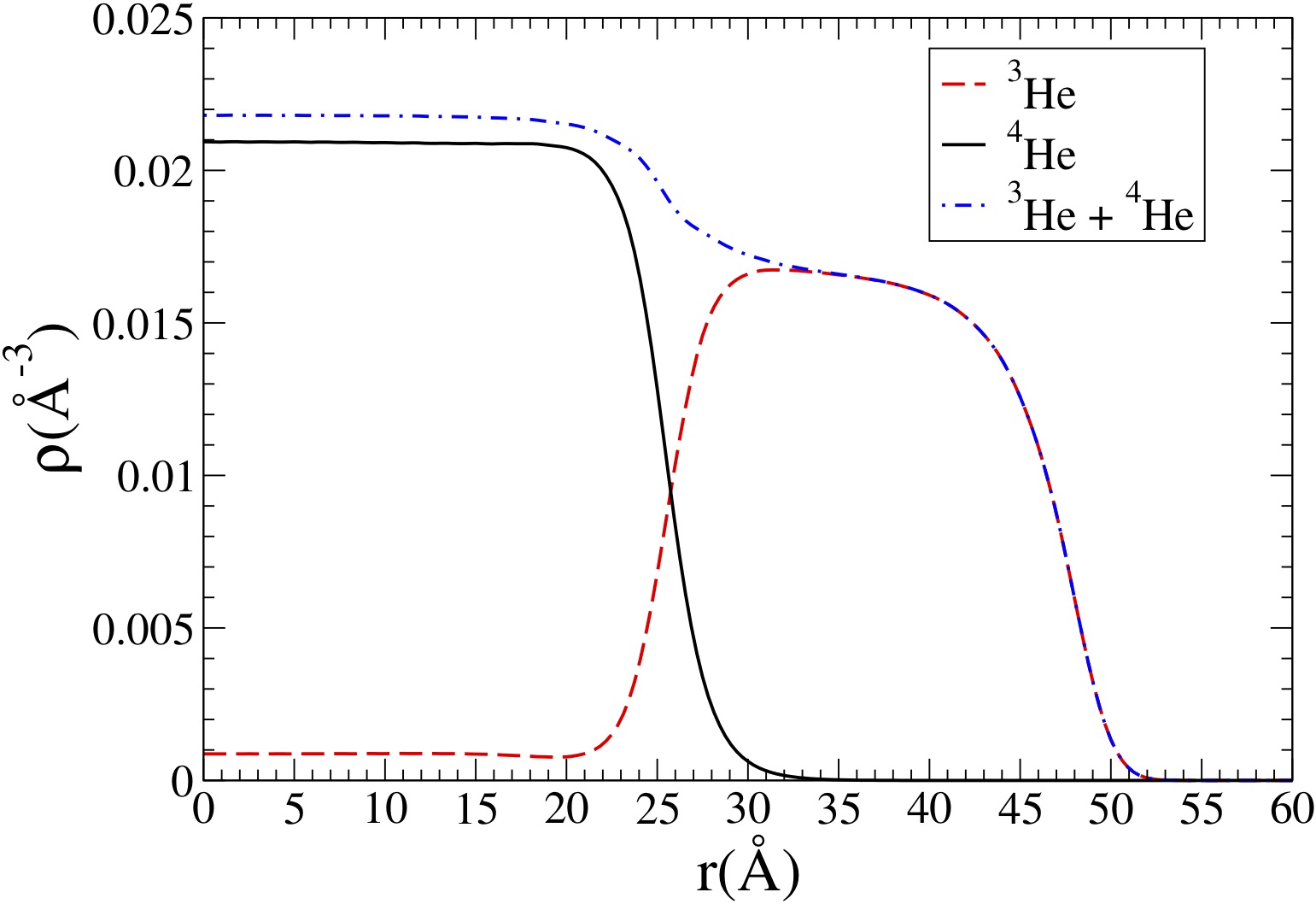}}
\caption{
Density profile of the spherical $^4$He$_{1500}$+$^3$He$_{6000}$ nanodroplet
as a function of the distance to the droplet center.
}
\label{figS1}
\end{figure*}
\begin{figure*}[!]
\centerline{\includegraphics[width=0.8\linewidth,clip]{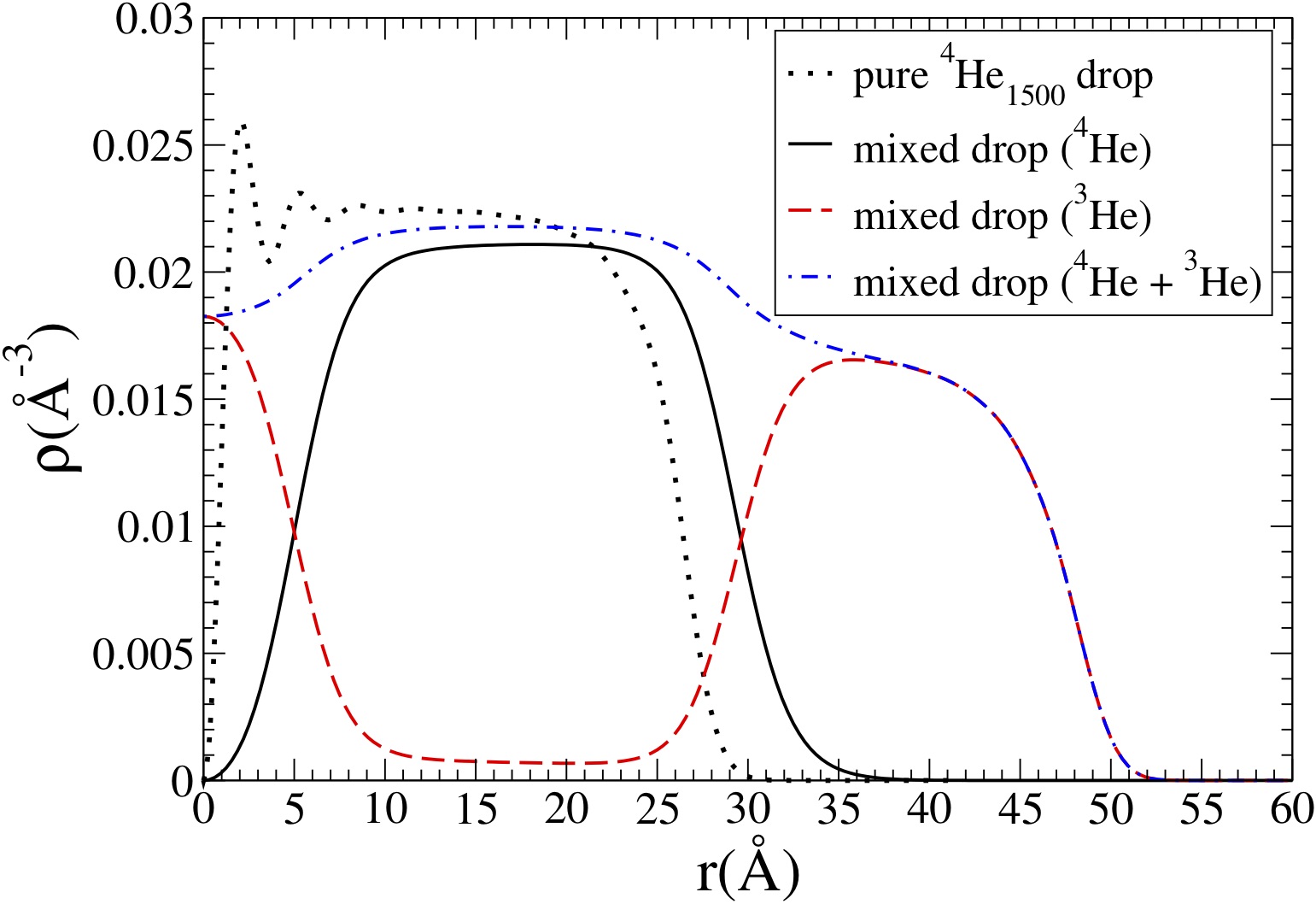}}
\caption{
Comparison between the density profile of the pure $^4$He$_{1500}$ droplet and that
of the  $^4$He core of the $^4$He$_{1500}$+$^3$He$_{6000}$ mixed droplet. Both nanodroplets host a vortex line.
In the later case the density profile of the $^3$He component is also shown.
Densities are shown along the radial direction within the symmetry plane perpendicular
to the vortex line.
}
\label{figS2}
\end{figure*}
\begin{figure*}[!]
\centerline{
\includegraphics[width=0.8\linewidth,clip]{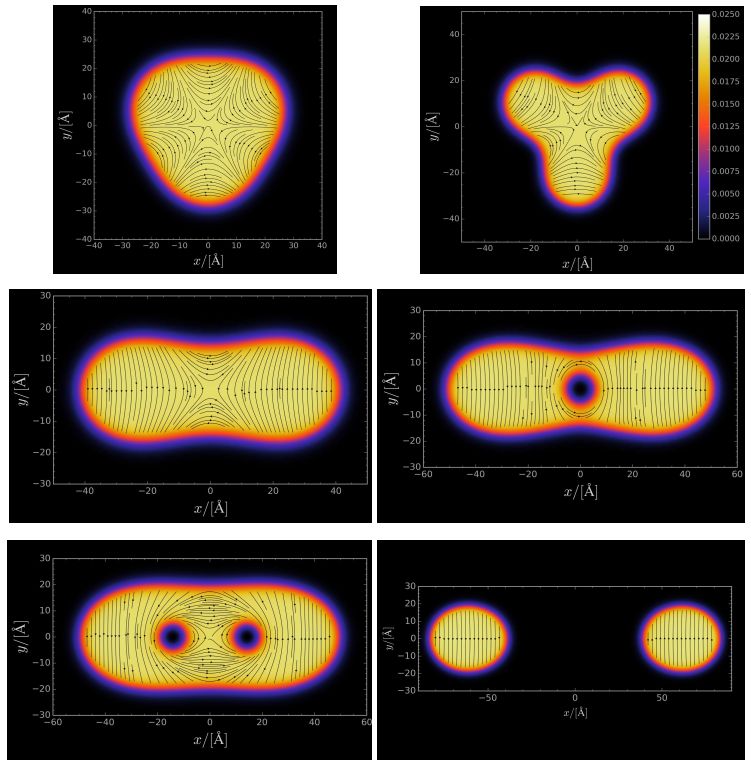}
}
\caption{
Density of the $^4$He core of the $^4$He$_{1500}$+$^3$He$_{6000}$ mixed nanodroplet
on the symmetry plane  perpendicular to the rotational axis. The streamlines of the
superfluid are superposed. Top: triangle-like and 3-lobed configurations at ${\cal L}=$ 3.64 and 4, respectively.
Middle: 
vortex-free and one-vortex prolate configurations at ${\cal L}=$ 3.5 and 4, respectively.
Bottom: two-vortex and fissioned  prolate configurations at ${\cal L}=$ 4 and 6, respectively.
The figures show streamlines arising from the presence of vortices (streamlines that wrap around the vortex cores) and others arising from 
the presence of capillary waves (streamlines that terminate at the droplet surface). They are associated to very different --but both irrotational--
 velocity fields.
 Notice the different length scales. The color bar shows the atom density in units of \AA$^{-3}$ and is common to all panels.
}
\label{S3}
\end{figure*}

\end{document}